\documentclass[amsmath,superscriptaddress,reprint,aps]{revtex4-2}
\usepackage{graphicx}
\usepackage{dcolumn}
\usepackage{bm}
\usepackage{hyperref}
\usepackage{siunitx}
\usepackage{color, soul}

\setstcolor{red}

\begin{document}
	
	
	\title{Architecture for coherent dual-comb spectroscopy and low-noise photonic microwave generation using mechanically actuated soliton microcombs}

      \author{Tatsuki~Murakami}
      	\affiliation{Department of Physics, Faculty of Science and Technology, Keio University, Yokohama, 223-8522, Japan}
        
     \author{Koshiro~Wada}
	\affiliation{Department of Electronics and Electrical Engineering, Faculty of Science and Technology, Keio University, Yokohama, 223-8522, Japan}

         \author{Soma~Kogure}
	\affiliation{Department of Electronics and Electrical Engineering, Faculty of Science and Technology, Keio University, Yokohama, 223-8522, Japan}

             \author{Ryomei~Takabayashi}
	\affiliation{Department of Electronics and Electrical Engineering, Faculty of Science and Technology, Keio University, Yokohama, 223-8522, Japan}

     \author{Liu~Yang}
      	\affiliation{Department of Physics, Faculty of Science and Technology, Keio University, Yokohama, 223-8522, Japan}
	\affiliation{Department of Electronics and Electrical Engineering, Faculty of Science and Technology, Keio University, Yokohama, 223-8522, Japan}
    \affiliation{Electronic Materials Research Laboratory, Key Laboratory of the Ministry of Education $\&$ International Center for Dielectric Research, School of Electronic Science and Engineering, Faculty of Electronic and Information Engineering, Xi’an Jiaotong University, Xi’an, 710049, China}
 
      \author{Riku~Shibata}
      	\affiliation{Department of Physics, Faculty of Science and Technology, Keio University, Yokohama, 223-8522, Japan}

     \author{Hajime~Kumazaki}
	\affiliation{Department of Physics, Faculty of Science and Technology, Keio University, Yokohama, 223-8522, Japan}
    \affiliation{Department of Electronics and Electrical Engineering, Faculty of Science and Technology, Keio University, Yokohama, 223-8522, Japan}

      \author{Shinichi~Watanabe}
      	\affiliation{Department of Physics, Faculty of Science and Technology, Keio University, Yokohama, 223-8522, Japan}

           \author{Atsushi~Ishizawa}
	\affiliation{College of Industrial Technology, Nihon University, Narashino, Chiba
275-8575, Japan}
        
    \author{Takasumi~Tanabe}
	\affiliation{Department of Electronics and Electrical Engineering, Faculty of Science and Technology, Keio University, Yokohama, 223-8522, Japan}

	\author{Shun~Fujii}
	\email[Corresponding author. ]{shun.fujii@phys.keio.ac.jp}
 	\affiliation{Department of Physics, Faculty of Science and Technology, Keio University, Yokohama, 223-8522, Japan}
	
	
\begin{abstract}
Dissipative Kerr soliton microcombs have inspired various intriguing applications such as spectroscopy, ranging, telecommunication, and high purity microwave generation. Mechanically actuated soliton microcombs provide enhanced controllability and flexibility for Kerr solitons, thus enabling technological progress to be made on such practical applications. Here, we present architectures for coherent dual-comb techniques and ultralow-noise microwave generation by exploiting the mechanical actuation of ultrahigh-Q crystalline microresonators. By unifying a pump laser, we demonstrate highly coherent dual-soliton combs using distinct resonators with slightly different repetition rates. We also report significant phase noise reduction achieved by directly generating Kerr solitons from a sub-Hz linewidth ultrastable laser. This study paves the way for further advancements in a wide variety of applications based on Kerr soliton microcombs.

\end{abstract}

\maketitle

Microresonator-based optical frequency combs, now known as microcombs, have various applications thanks to their fascinating features including compactness, high-repetition rate, low-power operation, and a broadband spectrum covering up to one-octave~\cite{Del’Haye2007, Karpov2019, Sayson2019}. Specifically, the high repetition rates of microcombs ranging from gigahertz to several hundreds of gigahertz have greatly accelerated the acquisition speed of dual-comb spectroscopy~\cite{Suh600, Bao2019a, Pavlov2017, Lucas2018, Bao2021, Weng2019, Yang2019} and ranging~\cite{Suh884, Lukashchuk2022, Wang2020}. 
Dissipative Kerr soliton (DKS) states provide a mode-locked pulse stream with a femtosecond pulse width alongside a broadband optical spectrum sustained by the double balance of dispersion and nonlinearity as well as parametric gain and cavity loss. Such unique features of DKSs have been utilized in dual-comb technologies to achieve accuracy, a high acquisition rate, and a small footprint simultaneously. To realize dual-DKS combs, we need twin microresonators with slightly different mode spacings, but achieving mutual coherence between two DKS sources is a challenge in this system~\cite{Suh600, Bao2019a, Pavlov2017}. As an alternative approach, counter-propagating~\cite{Yang2017, Joshi2018, Qureshi2023} and spatial multiplexing DKSs~\cite{Lucas2018, PhysRevLett.130.153802, Qureshi2023} have been employed to ensure coherence by unifying a pump laser and a resonator. Although these approaches have several advantages over the twin-resonator configuration, they unfortunately sacrifice flexibility in dual-combs such as that related to frequency differences in repetition rates and carrier offsets that are essential if we are to fully exploit dual-comb techniques~\cite{Lucas2018, Bao2021, Weng2019, Yang2019}.

\begin{figure}[b!]
\centering\includegraphics{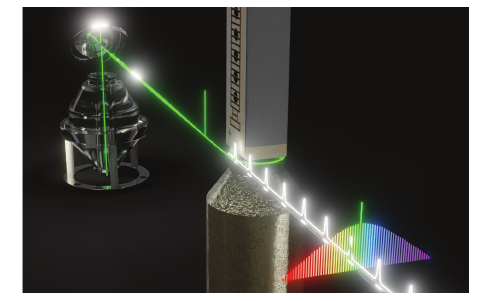}
\caption{\label{Fig_schematics} Conceptual schematics of mechanically actuated dissipative Kerr soliton (DKS) microcombs in a crystalline microresonator pumped by an ultra-stable laser.}
\end{figure}


Apart from dual-comb applications, the synthesis of ultralow-noise microwaves is of significant importance in metrology, radar, and satellite communications~\cite{Shao2024}. A promising way to produce low-noise microwaves is to use photonic systems, i.e., photonic microwave generation~\cite{Liu2020NP, PhysRevLett.122.013902}. Since optical frequency combs play an essential role as a gear between the optical and microwave frequency domains, they have been used to generate pure microwave signals over the past few decades~\cite{Diddams2020}. Meanwhile, microcombs are suitable form factors for directly producing high-frequency microwaves above 10~GHz as the repetition rates naturally fall within the range. One of the key elements in optical-to-microwave frequency division is an ultrastable laser (USL). A USL boasts unprecedented frequency stability and phase-noise performance via tight frequency locking to a stable high-finesse, ultralow expansion (ULE) cavity~\cite{Xie2017, Lucas2020, Qu2024}. Given that a microcomb is parametrically seeded from a continuous-wave (CW) laser, thus inheriting the linewidth and noise from the pump laser~\cite{Lei2022}, the royal road to photonic microwave synthesis is the use of the USL as the pump light to generate microcombs.

In this Letter, we present architectures for coherent dual-comb spectroscopy and photonic microwave generation by using mechanically actuated Kerr soliton microcombs. Soliton generation enabled by rapid resonance control allows us to obtain mutually coherent dual-combs as well as Kerr microcombs directly pumped by a sub-hertz linewidth ultrastable laser. Mechanical actuation allows us not only to extend the flexibility of the repetition frequencies in dual-DKS combs while maintaining the mutual coherence but also to enhance the potential ability of DKSs in the photonic synthesis of ultralow-noise microwaves. This study also envisions practical applications and deployment that fully exploit the features of Kerr microcombs while eliminating the system complexity involved in soliton generation and stabilization.

\begin{figure}[t!]
\centering\includegraphics{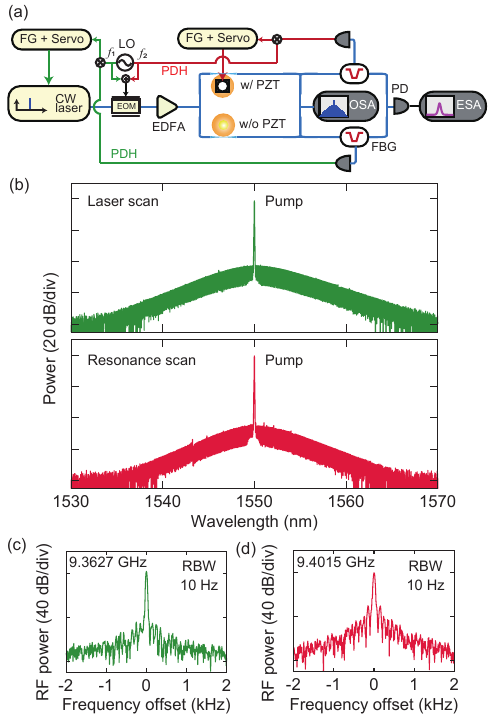}
\caption{\label{Fig_dualcomb} (a) Experimental setup for dual-DKS comb generation. FG, function generator; LO, local oscillator; PDH, Pound-Drever-Hall method; EOM, electro-optic modulator; EDFA, erbium-doped fiber amplifier; PZT, piezoelectric transducer; FBG, fiber Bragg grating filter; PD, photodetector; OSA, optical spectrum analyzer; ESA, electrical spectrum analyzer. The LO generates two-tone frequencies ($f_1 = 4.3$~MHz and $f_2 = 3.0$~MHz), which are used for Pound-Drever-Hall (PDH) locking. (b) Optical spectra of single DKS microcombs generated with laser scanning (green) and resonance scanning (red). The pump-detuning frequencies are 4.3~MHz and 3.0~MHz, respectively. (c, d) Soliton beat-note spectra representing the repetition rates.}
\end{figure}

First, we report mutually coherent dual combs with two individual magnesium fluoride ($\mathrm{MgF_2}$) crystalline microresonators. The details of direct soliton generation via mechanical actuation are described in Ref.~\cite{Fujii2024}. Mechanical shaping, polishing, and the careful treatment of resonator surfaces allow a significant Q-factor enhancement of up to $\sim10^9$.  The experimental setup is shown in Fig.~\ref{Fig_dualcomb}(a). One resonator is equipped with a lead-zirconate titanate (PZT) element to access a soliton state and ensure subsequent stabilization via the rapid control of the pump resonance frequency. We first generate a soliton microcomb with conventional pump laser scanning~\cite{Herr2014,Fujii2023} in a bare resonator. The pump laser is equally split after amplification of up to $\sim$1~W by using an erbium-doped fiber amplifier (EDFA). A portion of the transmitted power is used for Pound-Drever-Hall (PDH) locking~\cite{Drever1983,Fujii2023,PhysRevLett.121.063902} to stabilize the soliton state, where the feedback signal is fed to the pump laser frequency. We then drive the counterpart resonator to generate a soliton microcomb by mechanical actuation, and we also apply the PDH feedback through the same mechanical channel. It should be noted that thermo-electric cooled (TEC) devices enable wide-range resonance tuning with an efficiency of $\sim$1.8~GHz/K~\cite{Fujii2023}, making it possible to align desired soliton modes around a fixed pump wavelength. With this approach, two individual resonators now share a pump laser, an optical amplifier, and even an optical modulator to initiate and stabilize dual-combs. Importantly, mutual coherence is ensured because a single pump laser simultaneously excites soliton combs in two distinct resonators, which is the key to achieving coherent dual comb applications.

Figure~\ref{Fig_dualcomb}(b) shows optical spectra of dual soliton combs, where the solitons share the same pump laser.  The repetition rates ($f_\mathrm{rep}$) of each comb are monitored with a high-speed photodetector and an electrical spectrum analyzer (Fig.~\ref{Fig_dualcomb}(c) and \ref{Fig_dualcomb}(d)), they are 9.36267~GHz and 9.40149~GHz, respectively. Here, a mechanically actuated comb shows a slightly higher repetition rate, and the difference in repetition rates ($\Delta f_\mathrm{rep}$) corresponds to 38.82~MHz. This result indicates a compression factor $m$ of $f_\mathrm{rep}/\Delta f_\mathrm{rep}\approx 2.4\times10^2$ and the measurable bandwidth without spectral aliasing is  $f_\mathrm{rep}^2/(2\Delta f_\mathrm{rep}) \approx 1.1$~THz.   

\begin{figure}[t!]
\centering\includegraphics{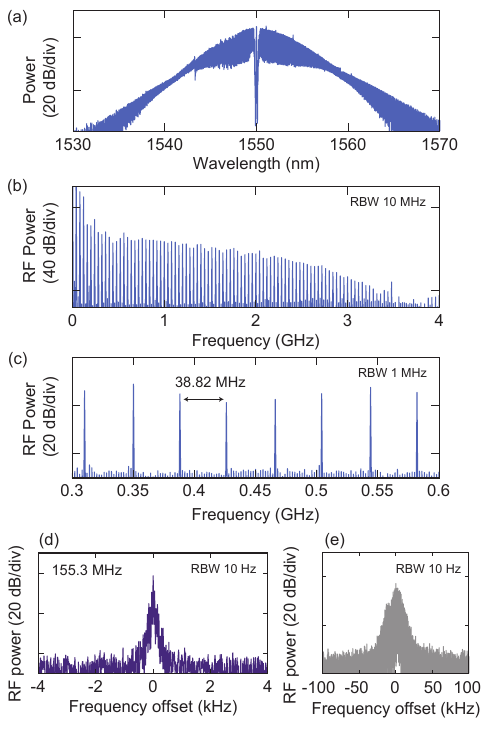}
\caption{\label{Fig_rfcomb}
(a) Combined optical spectrum of dual soliton combs. The residual pump light is filtered out by using an FBG filter. (b,c) Electrical spectrum obtained by photo detection of heterodyned DKSs. The frequency intervals correspond to the difference between soliton repetition rates. (d) Magnified view of 4th RF comb (155.3~MHz), which represents a very narrow linewidth comparable to the fundamental soliton beat note. (e) Extracted RF comb when distinct pump lasers are used to generate dual soliton combs. The broad linewidth (3-dB linewidth of $\sim$10~kHz) is mainly attributed to the relative frequency stability of the two pump lasers.}
\end{figure}

Two individual soliton combs are combined by using a 50$\times$ 50 fiber coupler after they have passed through a fiber-Bragg grating filter to eliminate any residual pump light.  An interference optical spectrum is presented in Fig.~\ref{Fig_rfcomb}(a). We observe an interferogram spectrum in the electrical domain, i.e., RF comb, as a result of down-conversion from the optical domain (Fig.~\ref{Fig_rfcomb}(b) and \ref{Fig_rfcomb}(c)). The RF comb lines are equidistantly separated by 38.82~MHz with a signal-to-noise ratio of 30~dB. Here, dual combs share the same pump laser and the comb lines on both sides with respect to the pump frequency will have equal heterodyne signals, making them indistinguishable in the RF domain. To avoid this limitation, a frequency shifter (e.g., an acousto-optic modulator) can be added on one side of dual combs. A close-up view of one of the RF combs (Fig.~\ref{Fig_rfcomb}(d)) indicates a significant narrow linewidth, which is comparable to the beat-note linewidth of the soliton repetition rate. In contrast, the RF comb exhibits a much broader linewidth when distinct pump lasers are used to generate dual combs due to the absence of coherence and relative frequency drift as shown in Fig.~\ref{Fig_rfcomb}(e). This is clear evidence that the coherence of the dual combs would degenerate when we use different pump lasers. Thanks to the use of a single-pump, two-cavity configuration, we can enhance the flexibility of microresonator-based dual comb techniques, for instance, the controllability of $\Delta f_\mathrm{rep}$ while maintaining both mutual coherence and system simplicity. In principle, the difference in repetition rates can be precisely controlled by employing dispersion engineering~\cite{FujiiTanabe+2020+1087+1104} and ultraprecision machining~\cite{Fujii:20} and aligned by thermal tuning with a rate of ${d f_\mathrm{rep}}/{dT}\approx f_\mathrm{rep}\times10^{-5}$.

\begin{figure}[ht!]
\centering\includegraphics{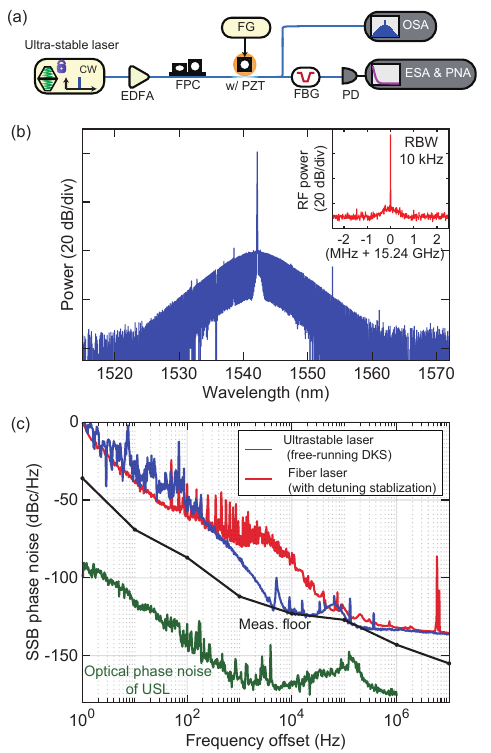}
\caption{\label{Fig_usl} (a) Experimental setup for photonic microwave generation. FPC, fiber polarization controller; PNA, phase noise analyzer. (b) Optical spectrum of a single soliton microcomb directly pumped by an USL. The inset shows a soliton beat note, yielding a repetition rate of 15.24~GHz. (c) Single-sideband (SSB) phase noise spectrum of the fundamental repetition rate of a single soliton. The solid blue line shows the phase noise PSD of a microwave obtained with the USL-pumped, free-running DKS. The solid red line shows the phase noise of the DKS obtained with laser scanning, where PDH detuning stabilization is applied. The optical phase noise scaled to 15.24~GHz is also indicated by the solid green line. The black line is the measurement floor of the phase-noise analyzer.}
\end{figure}

Next, we present a low-noise DKS directly pumped by a USL.  Earlier studies have revealed that soliton microcombs receive several noise contributions from including the phase and intensity noise of the pump laser, shot noise, and thermo-refractive noise~\cite{Yang2021,Lei2022}. Importantly, each comb line of the DKS directly inherits the noise properties of the pump laser owing to coherent parametric processes, and therefore a promising way of reducing the comb noise is to exploit an ultra-stable, ultralow-noise CW laser as the pump laser. Figure~\ref{Fig_usl}(a) shows the experimental setup. Here, we use an ultra-stable reference laser (Menlo Systems, ORS1500) with a wavelength of 1542 nm, where a CW laser is tightly locked to a high-finesse cavity made of ultra-low expansion (ULE)
glass, ensuring an ultimate low-noise pump source with a linewidth of less than 1~Hz. Although the frequency of the USL cannot be tuned, mechanical actuation readily allows the generation of soliton microcombs. Figure~\ref{Fig_usl}(b) shows the observed spectrum of a single soliton state followed by the optical amplification and polarization adjustment of the pump laser (Fig.~\ref{Fig_usl}(a)). The inset shows the beat-note spectrum, yielding a repetition rate of 15.24 GHz. It should be noted that no active feedback such as detuning stabilization is performed in this experiment.

Of particular interest is the phase noise of the free-running soliton repetition rate because the USL is expected to exhibit an outstanding phase noise property compared with those of commonly-used pump sources such as an external cavity diode laser or a fiber laser. Figure~\ref{Fig_usl}(c) shows the phase noise power spectrum density (PSD) of the fundamental repetition rate. At offset frequencies of 100~Hz to 100~kHz, the USL-pumped DKS exhibits greatly suppressed phase-noise performance compared with the DKS pumped by a fiber laser (NKT Photonics, Koheras E15). It is noteworthy that the phase noise decreases by up to approximately 50 dB at around 5~kHz, reaching a lower limit imposed by the measurement floor. The PSD has a gentle peak at 100~kHz, where a similar peak can be found in the optical phase noise of the USL due to the servo control bandwidth, i.e., servo bump. This indicates that the soliton microcomb inherits the characteristics of the USL. It should be noted that 
large variations at low offset frequencies ($<100$~Hz) arise from cavity temperature change and subsequent detuning fluctuation, which can be compensated for by applying pump-detuning stabilization. We confirmed that the variation in the PSD is clearly reduced when the PDH feedback is supplied to the pump laser frequency below its control bandwidth ($\sim$ a few kHz) as shown by the solid red line in Fig.~\ref{Fig_usl}(c). In short, the phase noise of a soliton microcomb is significantly suppressed thanks to the superior frequency stability inherent to the USL. Further improvements in the SSB phase noise can be expected if we operate the DKS at a quiet point~\cite{Lucas2020,Yao2022} with pump-detuning stabilization.

In conclusion, we have demonstrated two architectures using mechanically actuated Kerr soliton microcombs in ultrahigh-Q crystalline microresonators. Dual-DKS combs were generated in twin microresonators with slightly different repetition rates from a single pump laser, which ensured mutual coherence between two individual DKSs. The interference spectrum in the radio-frequency domain, namely the RF comb, exhibits a narrow linewidth feature with an interval of 38.82~MHz that corresponds to the difference in the repetition frequencies. These results reinforce the flexibility of the dual comb parameters such as $\Delta f_\mathrm{rep}$ while eliminating the system's complexity. This study also offers several advantages as regards dual-comb spectroscopy and distance measurements, where coherent averaging is the key to enhancing the signal-to-noise ratio. A proof-of-concept experiment for ultralow-noise microwave generation was also demonstrated. Direct soliton generation from a frequency-fixed, sub-Hz linewidth USL provides us with an extremely simple scheme with which to reduce the phase-noise of soliton microcombs. Combined with sophisticated frequency stabilization methods proposed elsewhere~\cite{Lucas2020, Wildi2024, Wu2023}, mechanically actuated DKSs could lead to innovative improvements in relation to photonic microwave generation, optical frequency division and optical clock technologies.


\section*{Acknowledgments}
This work was supported by JSPS KAKENHI (JP24K17624, JP24K01388, JP24K21743, JP23K26165); Adaptable and Seamless Technology transfer Program through Target-driven R\&D (A-STEP) from Japan Science and Technology Agency (JST) (JPMJTR23RF); Ministry of Education, Culture, Sports, Science and Technology (JPMXS0118067246); Keio University Program for the Advancement of Next Generation Research Projects; T. Murakami acknowledges JST SPRING (JPMJSP2123). S. Fujii acknowledges the support from Mizuho Foundation for the Promotion of Sciences; The Murata Science Foundation. We thank Katsuya Oguri at NTT Basic Research Laboratories, Koki Yoshida and Yugo Kikkawa at Tokyo Denki University for support in experimental setup. We are grateful for the technical support provided by the Manufacturing Center at Keio University.

\section*{Disclosures}
The authors declare no conflicts of interest.

\section*{Data availability} Data underlying the results presented in this paper are obtained from the authors upon reasonable request.


\bibliography{Soliton_20241117}


\end{document}